\documentclass[12pt,journal,onecolumn]{IEEEtran}
\usepackage[doublespacing]{setspace}
\usepackage{lineno}
\usepackage{amsmath,amsfonts}
\usepackage{algorithmic}
\usepackage{algorithm}
\usepackage{array}
\usepackage[caption=false,font=normalsize,labelfont=sf,textfont=sf]{subfig}
\usepackage{textcomp}
\usepackage{stfloats}
\usepackage{url}
\usepackage{verbatim}
\usepackage{graphicx}
\usepackage{cite}
\usepackage{color}
\usepackage{multirow}
\usepackage{booktabs}
\usepackage{tabularx}
\usepackage{array}
\usepackage{bm}
\usepackage{gensymb}
\begin{document}
\title{Normal mode parameters estimation by a VLA in single-shooting}

\author{Xiaolei Li, Pengyu Wang, Wenhua Song*, Yangjin Xu, Wei Gao
        % <-this % stops a space
%\thanks{This paper was produced by the IEEE Publication Technology Group. They are in Piscataway, NJ.}% <-this % stops a space
%\thanks{Manuscript received April 19, 2021; revised August 16, 2021.}
\thanks{This work was supported in part by the National Natural Science Foundation of China under Grants 12474444, 12104429, 52071309, 12004359.}
\thanks{The authors would like to thank Professor Ning Wang for his
valuable suggestions.}}

\markboth{Journal of \LaTeX\ Class Files,~Vol.~14, No.~8, August~2021}%
{Shell \MakeLowercase{\textit{et al.}}: A Sample Article Using IEEEtran.cls for IEEE Journals}


\maketitle

\begin{abstract}
This paper proposes an orthogonality-constrained modal search (OCMS) method for estimating modal wavenumbers and modal depth functions using a vertical linear array (VLA). 
Under the assumption of a known sound speed profile, OCMS leverages \textcolor{red}{ the orthogonality of distinct modal depth functions} to extract both the modal depth functions and their corresponding wavenumbers, even when the VLA and a monochromatic sound source remain stationary.The performance of OCMS is evaluated through numerical simulations under varying signal-to-noise ratios (SNRs), different VLA apertures, varying numbers of VLA elements, VLA tilt and sound speed profile (SSP) uncertainty. The results demonstrate that OCMS is robust against noise, VLA aperture variations, and \textcolor{red}{changes in the number of VLA elements}, meanwhile, the algorithm maintains reliable performance when SSP uncertainty $<$ 1 m$/$s and VLA tilt angle $<5\degree$. Furthermore, the effectiveness of OCMS is validated using SwellEx96 experimental data. The relative error between the modal wavenumbers derived from experimental data and those computed via Kraken is \textcolor{red}{on the order of $10^{-4}$}.
\end{abstract}


\section{Introduction}
%\linenumbers
In shallow water, sound propagation can be characterized by normal mode theory.
Modal wavenumbers (MWs) and mode depth functions (MDFs) are fundamental parameters in this framework, playing a critical role in applications such as source localization \cite{ref1,ref2,ref3},
geoacoustic inversion\cite{ref4,ref5,ref6,ref23}, sound speed profile (SSP) inversion \cite{ref7}, etc.
The environmental mismatch problem encountered in matched modal processing (MMP) \cite{ref1,ref2} is primarily due to the discrepancies between the estimated MWs, MDFs, and their corresponding true values. Due to the importance of normal mode parameters in underwater acoustic applications,
\textcolor{red}{a lot of research has been done} for MWs and MDFs inversion \cite{ref8}-\cite{ref32}.

\par The estimation of MWs is typically performed using data from a horizontal linear array (HLA) or a synthetic horizontal array (SHA) generated by a moving source in a homogeneous waveguide. Classical estimation methods are mainly based on the Hankel transform approach \cite{ref8,ref9,ref10}.
Accurate estimation of MWs via the Hankel transform requires the HLA or SHA aperture to exceed the maximum modal cycle distance, a condition that is often impractical in real-world applications. To reduce the required HLA aperture, various high-resolution (HR) methods have been developed for MW estimation, including the autoregressive model \cite{ref11,ref12} , Prony methods \cite{ref13,ref14,ref15},
matrix pencil (MP) methods \cite{ref16}, subspace separation
methods \cite{ref17,ref18}  and methods based on compressed sensing \cite{ref19,ref20,ref21}.
\textcolor{red}{However, the above HR methods exhibit significant noise sensitivity, }ultimately limiting their ability to achieve a substantial reduction in the HLA aperture in practical scenarios.
To further reduce the array aperture in the horizontal range direction during MW estimation,\textcolor{red}{ conventional MUSIC and MP algorithms }designed for HLA configurations can be extended to combined VLA-SHA configurations, specifically incorporating a fixed VLA with a moving source\cite{ref22}.
 The aforementioned MW estimation methods generally require a precise measurement of either the moving source distance or the hydrophone positions within a horizontal range sufficient to resolve \textcolor{red}{ the minimum separation between adjacent MWs}.
 However, obtaining accurate range information for the VLA-SHA configuration is not always feasible. 
 On the one hand, \textcolor{red}{global navigation satellite systems (GNSS) have errors of certain level, even though} some reports indicate that GNSS can achieve centimeter-level positioning accuracy \cite{ref24,ref25}.
However, wave and current effects on both the receiver array and the sound source are unavoidable, introducing additional unpredictable random errors in the VLA-SHA range estimation.

\par The estimation of normal mode parameters using a VLA circumvents the inaccuracies associated with sound source-to-VLA range measurements. Theoretical studies have shown that MDFs can be successfully extracted through subspace decomposition methods when three conditions are satisfied: (1) adequate depth sampling of propagating modes by VLA receivers, (2) sufficient source-to-VLA range sampling, and (3) an appropriate range interval traversed by the sound source \cite{ref26}. The MWs can subsequently be estimated from the extracted mode depth functions (MDFs) given the knowledge of the sound speed profile at the VLA location\cite{ref27,ref28}.
  Although exact measurements of the source-to-VLA range are not required in the methods described in \cite{ref27} and \cite{ref28}, the moving source must traverse a range interval substantially exceeding the maximum modal cycle distance. Furthermore, the methods proposed in \cite{ref27} and \cite{ref28} require the deployment of a densely sampled VLA that spans the entire water column to acquire orthogonal mode depth functions (MDFs), a requirement that significantly constrains their practical applicability. 
   In 2022, F. Hunter Akins and W. A. Kuperman introduced the modal-MUSIC method, which enables extraction of normal modes from a moving source at unknown ranges using a partially-spanning vertical linear array (VLA), given prior knowledge of the water column's SSP \cite{ref36}.
   In 2023, we proposed the combining physical informed neural network (CPINN) method for estimating normal mode parameters \cite{ref29}.
  The CPINN estimates modal wavenumbers by introducing modal depth function proxies constrained to satisfy their corresponding differential equations. This approach eliminates the requirement for range-direction aperture exceeding the maximum modal cycle distance and enables modal wavenumber estimation without full water column spanning by the VLA. Furthermore, the CPINN remains applicable to range-dependent waveguides, where it can estimate modal wavenumbers at the VLA position. \textcolor{red}{All aforementioned VLA-based methods for normal mode parameter estimation require varying the source-to-VLA range during measurements.} For fixed source-range configurations, normal mode parameter estimation using a vertical linear array (VLA) can alternatively be accomplished through either: (1) depth variation of the sound source \cite{ref30}, or (2) use of broadband acoustic signals \cite{ref31,ref32}. 
If we define "single-shooting" as a single-frequency measurement conducted after determining the relative source-VLA position, then all aforementioned methods share the common requirement of multiple-shooting measurements. The capability to estimate normal mode parameters through single-shooting VLA measurements would offer significant operational advantages. However, to date, no such methodology has been reported in the literature to our knowledge. 

\par This paper presents the orthogonality constrained modal search (OCMS) method for estimating normal mode parameters using a VLA in single-shooting configurations. \textcolor{red}{Note that with known water column SSP,} the governing differential equations (or difference equations) for mode MDFs establish a fundamental relationship between the MWs and MDFs \cite{ref27,ref28}.
Additionally, different MDFs are approximately orthogonal to each other in water column. Based on that fact, by determining the MW of a specific normal mode, one can determine the MDF of that mode, as well as the MWs and MDFs of other modes.
The OCMS method utilizes the inherent relationship between MWs and MDFs, combined with the orthogonality property among distinct MDFs, to reduce the multidimensional normal mode parameter estimation problem to a one-dimensional MW search problem. 
The OCMS method requires the VLA element count to be equal to or exceed the number of propagating modes received, but eliminates the conventional requirement for a full water column spanning by the VLA. The effectiveness of OCMS is verified by simulations and \textcolor{red}{experimental data in this paper}.

\par This paper is organized as follows. We will introduce the OCMS method in detail in Section II.
In Section III, the performance of normal mode parameter estimation by OCMS under different conditions will be analyzed by simulations. In Section IV, the effectiveness of the OCMS method will be verified by experimental data. Finally, a conclusion will be given in Section V.

\section{Theory of orthogonality constrained modal search method}

MWs and MDFs represent fundamental normal mode parameters that are intrinsically related through the following governing differential equation:
\begin{equation}
\label{eq1}
       \frac {d^2}{dz^2 } \phi_m (z,k_m )+(k^2 (z)-k_m^2 ) \phi_m (z,k_m )=0,	
\end{equation}
where $z$ is the depth coordinate, $k(z)=\omega/c(z)$ is the wavenumber, $\omega$ is the angular frequency,
 $c(z)$ is the SSP, $\phi_m (z,k_m )$ is the $m\mathrm{th}$ MDF, and $k_m$ is the corresponding MW.
 The term $k_m$ in $\phi_m (z,k_m )$ explicitly indicates the functional dependence of the MDF on its corresponding MW. \textcolor{red}{It is convenient for numerical calculations to modify the above differential equation to the following difference equation} \cite{ref28}
\begin{equation}
\begin{array}{l}
\phi_m(z+h,k_m )+\phi_m(z-h,k_m )\\
=2 \mathrm{cos} (\sqrt{(k^2 (z)-k_m^2 )}  h) \phi_m (z,k_m )
\end{array}
\label{eq2}
\end{equation}
\textcolor{red}{where the sound speed is assumed to be constant across the interval $[z-h, z+h]$ with $h$ representing the discretization step size.}
Based on Eq. (\ref{eq2}), one can use $k_m$ to estimate $\phi_m (z,k_m )$ by a shooting method with the shooting initial conditions
\begin{equation}
\label{eq3}
\begin{cases}
\phi_m (0,k_m )=0
 \\
\phi_m (h,k_m )=1
\end{cases}
\end{equation}
In Eq.(\ref{eq3}), the free boundary condition on the sea surface has been used.
\textcolor{red}{Although determining $M$ MDFs requires $M$ MWs in general, under the assumption that the MDFs in the water column are orthogonal to each other, all MDFs in the water column are determined when any MW is given, i.e, the number of free parameters is one.}
Here, the orthogonality of the MDFs in the water column can be expressed by the following equation:
\begin{equation}
\label{eq4}
\begin{array}{c}
\left\langle\psi_{m}\left(z, k_{m}\right), \psi_{n}\left(z, k_{n}\right)\right\rangle=\sum_{l=1}^{L} \psi_{m}\left(z_{l}, k_{m}\right) \psi_{n}^{*}\left(z_{l}, k_{n}\right) h \\
\approx \int_{0}^{H} \psi_{m}\left(z, k_{m}\right) \psi_{n}^{*}\left(z, k_{n}\right) d z \\
\approx \int_{0}^{\infty} \psi_{m}\left(z, k_{m}\right) \psi_{n}^{*}\left(z, k_{n}\right) d z=\delta_{m n},
\end{array}
\end{equation}
where $z_l=lh$, $h\ll1$,  $m,n\in \mathbb{N}^+$,$H\in[z_L,z_{L+1} )$ is the water depth,  $\delta_{mn}$ is the Kronecker delta function and
\begin{equation}
\label{eq5}
\psi_m (z,k_m )=\frac {(\phi_m (z,k_m ))}{\left \| \phi_m (z,k_m ) \right \|_2 }  .
\end{equation}
where
\begin{center}
$\left \| \phi_m (z,k_m ) \right \|_2 =\sqrt{\sum_{l=1}^{L} \phi_{m}\left(z_{l}, k_{m}\right) \phi_{n}^{*}\left(z_{l}, k_{n}\right) h}$.
\end{center}
The second ``$\approx$'' in Eq.(\ref{eq4}) highlights that strict orthogonality of the MDFs is only achieved when the depth domain spans $0\leq z \leq \infty$.
The validity of the MDF orthogonality assumption in the water column has been confirmed by successful MDF extraction \cite{ref26,ref27,ref28}.
\begin{algorithm}[H]
\caption{}\label{alg:alg1}
\begin{algorithmic}
\STATE
\STATE {\textbf{Input: }}$\xi_1,c(z), \xi_{min},\xi_{max},\Delta\xi,\alpha=1,\beta=2$
\STATE  \textbf{Compute} $\psi_1 (z_l,\xi_1 )$ with $l=1,2,\cdots,L$ by combing Eqs. (2), (3) and (5).
\STATE \textbf{While} $ \xi_{min}\leq\xi_1-\alpha\Delta\xi\leq\xi_{max}$
\STATE \hspace{0.5cm} \textbf{Compute} $  \psi(z_l,\xi_1+\alpha\Delta\xi)$ by combing Eqs. (2), (3) and (5) with $l=1,2,\cdots,L.$
\STATE \hspace{0.5cm} \textbf{Compute} $     c_\alpha=\left\langle\psi_1 (z_l,\xi_1 ), \psi(z_l,\xi_1+\alpha\Delta\xi)\right\rangle.$
\STATE \hspace{0.5cm} \textbf{If} $ 2<\alpha<[(\xi_{max}-\xi_{min})/\Delta\xi], c_{\alpha-2}<c_{\alpha-1}, c_\alpha<c_{\alpha-1}$ and $c_{\alpha-1}\ll1$
\STATE \hspace{1cm} $		\xi_\beta=\xi_1+(\alpha-1)\Delta\xi$
\STATE \hspace{1cm} $		\psi_\beta (z_l,\xi_\beta )=\psi(z_l,\xi_\beta )$
\STATE \hspace{1cm} $		\beta=\beta+1$
\STATE \hspace{0cm} $	\alpha=\alpha+1$
\STATE \hspace{0cm} $	M=\beta-1$
\STATE \textbf{Output: }$ \xi_m, \psi_m (z_l,\xi_m )$ with $m=1,2,\cdots,M$ and $l=1,2,\cdots,L.$

\end{algorithmic}
\label{alg1}
\end{algorithm}

Now we introduce the OCMS method. Given a MW $\xi_m$, 
a set of MDFs 
\[
\begin{array}{l}
\Phi(z,\xi_m )\\:= \{\psi_1 (z,\xi_1 ),\psi_2 (z,\xi_2 ),\cdots,\psi_m (z,\xi_m ),\cdots,\psi_M (z,\xi_M ) \}
\end{array}
\]
can be obtained by \textcolor{red} { combining Eqs (\ref{eq2}) to (\ref{eq4}). It is easy to show} 
\[
\Phi(z,\xi_1 )=\Phi(z,\xi_2 )=\cdots=\Phi(z,\xi_{M} ).
\]
\textcolor{red}{Given $\xi_1$ as the MW of the first normal mode, a computational procedure is provided to determine $\Phi(z,\xi_1 )$ as shown by Algorithm 1. } In Algorithm 1, we can choose $\xi_{min} \leq \omega/c_b$ with $c_b>\mathrm{max}(c(z))$ being the bottom sound speed and $\xi_{max}=\omega/\mathrm{min}(c(z))$. $\Delta\xi\in \mathbb{R}$ is the step size of the MW search, and ``$[a]$'' represents the integer part of the real number $a$ in Algorithm 1. The mode number associated with a given MW can be determined by counting the nodal points (zero crossings) in its corresponding mode MDF. 
\par For sufficiently distant sound sources, the acoustic field in the depth direction can be represented by a finite sum of \textcolor{red}{ propagating }normal modes: 
\begin{equation}
\label{eq7}
p(z)=\sum_{m=1}^{M} a_m \psi_m (z,k_m ) .
\end{equation}
where $M \in \mathbb{N}^+ $ denotes the number of propagating modes. The mode amplitude is given by
\begin{center}
$a_m= \frac{i}{4} S(\omega) \psi_m (z_s,k_m ) H _0 ^{(1)} (k_m r),$
\end{center}
where $i^2=-1$, $S(\omega)\in \mathbb{C} $ represents the source spectrum at angular frequency $\omega$, $z_s$ is the sound source depth, and $r$ denotes the horizontal distance between the source and the observer point.
 Under the assumption of the orthogonality of the MDFs in the water column, $\psi_m (z,k_m )\in \Phi(z,k_1 )$ holds for $m=1,2,\cdots,M$, while $\Phi(z,\xi_1 )\neq \Phi(z,k_1 )$ when $\xi_1\neq k_1$. This implies the following non-orthogonal condition: 
\begin{center}
$\langle\psi_n (z,\xi_n ),\psi_m (z,k_m )\rangle \neq 0,\qquad m,n \in \mathbb{N}.$
\end{center}

The Appendix clarifies that when $\xi_1\neq k_1$, the acoustic field $p(z)$ cannot be expressed as a finite linear combination of elements from $\Phi(z,\xi_1 )$. This demonstrates that only  $\Phi(z,k_1 )$ provides a suitable finite-dimensional representation for $p(z)$.  Consequently, through systematic exploration of potential $\xi_1$ values and their corresponding $\Phi(z,\xi_1 )$, the MWs and MDFs can be estimated by verifying whether the acoustic field $p(z)$ admits a finite linear representation using elements from $\Phi(z,\xi_1 )$. This is the basic idea of the OCMS method. In practical implementations, $p(z)$ is sampled through VLA measurements.
Let the depth coordinates of the VLA elements be represented by the vector $\mathbf{z}=[z_1,z_2,\cdots,z_N ]^T$, where $z_n$ denotes the depth of the $n\mathrm{th}$ element. The acoustic field in Eq. (\ref{eq7})  can then be expressed in matrix form as:
\begin{equation}
\label{eq9}
\mathbf{p}=\mathbf{\Psi}(k_1 )\mathbf{a}(k_1 ),
\end{equation}
 where 
 \[
 \mathbf{p}=[p(z_1 ),p(z_2 ),\cdots,p(z_N )]^\mathrm{T},
 \]
 \[
 \mathbf{\Psi}(k_1)=[\bm{\psi}_1 (k_1 ),\bm{\psi}_2 (k_2 ),\cdots,\bm{\psi}_M (k_M )],
 \]
\[
\bm{\psi}_m (k_m )=[\psi_m (z_1,k_m ),\psi_m (z_2,k_m ),\cdots,\psi_m (z_N,k_m )]^\mathrm{T} 
 \]
 with $ m=1,2,\cdots,M$, and $\mathbf{a}(k_1 )\in \mathbb{C}^{M\times1}$.
 
 \par First, \textcolor{red}{we examine the case where $M < N$, indicating that the VLA must contain more elements ($N$) than the number of observed propagating modes ($M$) .}
 Furthermore, the depth coordinate vector $\mathbf{z}$ should be selected to ensure the full column rank condition of the $\mathbf{\Psi}(k_1 )$.
 Since $p(z)$ admits finite linear representation exclusively through $\Phi(z,k_1 )$, it follows directly that $ ||\mathbf{p}-\mathbf{\Psi}(\xi_1 )\mathbf{a}(\xi _1 )||_2 >0 $ when $\xi_1\neq k_1$.  For the case where  $M\geq N$, there exists a coefficient vector  $\mathbf{a}(\xi _1 ) $ that exactly satisfies $ ||\mathbf{p}-\mathbf{\Psi}(\xi_1 )\mathbf{a}(\xi _1 )||_2 =0. $
Given the sparsity of received modes, we formulate the estimation problem as finding the sparsest coefficient vector  $\mathbf{a}(\xi _1 )$ that satisfies the exact representation condition $||\mathbf{p}-\mathbf{\Psi}(\xi_1 )\mathbf{a}(\xi _1)||_2=0 $. This yields the optimization problem: 
\begin{center}
 $\hat{k}_1=\mathop{arg}\limits_{\xi_1}\textbf{min} ||\mathbf{a}(\xi _1  )||_0$
\end{center}
		subject to
\begin{equation}
\label{eq10}
||\mathbf{p}-\mathbf{\Psi}(\xi_1 )\mathbf{a}(\xi_1 )||_2=0,
\end{equation}
where $ \hat{k}_1 $ denotes the estimated value of $k_1$. Eq. (\ref{eq10}) is a non-convex optimization problem, which is typically converted to a convex optimization problem by convex relaxation processing \cite{ref37}. 
Accounting for noise effects, we reformulate Eq. (\ref{eq10})  as the following $l_1\mathrm{-norm}$ optimization problem: 
\begin{center}
 $\hat{k}_1=\mathop{arg}\limits_{\xi_1}\textbf{min} ||\mathbf{a}(\xi _1  )||_1$
\end{center}
		subject to
\begin{equation}
\label{eq11}
||\mathbf{p}-\mathbf{\Psi}(\xi_1 )\mathbf{a}(\xi_1 )||_2<\varepsilon,
\end{equation}
where $\varepsilon>0$ can be determined by the signal-to-noise ratio. In this paper, \textcolor{red}{we choose $\varepsilon$ to be the noise energy.} The convex optimization problem in  Eq. (\ref{eq11}) can be \textcolor{red}{ solved by CVX},
a package for specifying and solving convex programs \cite{ref38,ref39}. 
By integrating Eq. (\ref{eq11}) with Algorithm 1, we develop a complete framework for estimating both MWs and MDFs. Algorithm 2 presents the detailed computational procedure for this joint parameter estimation.
 In Algorithm 2, the first MW search range is bounded by $\xi_{max}\in \mathbb{R}$ and $\xi_{min}\in \mathbb{R} $,  representing the upper and lower limits, respectively, while $\Delta \xi \in \mathbb{R}$ denotes the search step size.  Notably, while Algorithm 1 utilizes the orthogonality property of MDFs, the VLA implementation does not require full coverage of the water column, demonstrating that the OCMS method can operate without a VLA that covers the entire water column. 
Furthermore, Algorithm 2 demonstrates that the OCMS method achieves simultaneous estimation of all MWs and MDFs from a single snapshot. \textcolor{red}{Unlike conventional modal parameter estimation methods that only resolve parameters of excited modes, the OCMS method fundamentally provides complete modal parameter estimation in one computational process.}
 Because CVX returns a meaningful solution only when$ ||\mathbf{p}-\mathbf{\Psi}(\xi_1 )\mathbf{a}(\xi _1  )||_2 <\varepsilon$ can be satisfied,
 Algorithm 2 is also suitable for the case $M<L $.
 
 \par In the following, the performance of the OCMS method will be validated by simulations and \textcolor{red}{experimental data.}
 \begin{algorithm}[H]
\caption{}\label{alg:alg2}
\begin{algorithmic}
\STATE
\STATE {\textbf{Input: }}$\mathbf{p},\xi_1, \xi_{min},\xi_{max},\Delta\xi,l=1$
\STATE \textbf{While} $ \xi_{min}\leq\xi_1 \leq\xi_{max}$
\STATE \hspace{0.5cm} \textbf{Obtain} $ \mathbf{\Psi}(\xi_1 ) $ by the Algorithm 1
\STATE \hspace{0.5cm} \textbf{Solve}   Eq. (9) by CVX
\STATE \hspace{0.5cm} $ e(l)=\|\mathbf{a}(\xi_1 )\|_1 $
\STATE \hspace{0.5cm} \textbf{If} $ l>1 $ and $e(l)<e(l-1)$
\STATE \hspace{1cm} $	\hat{k}_m=\xi_m$ with $m=1,2,\cdots,M$
\STATE \hspace{0.5cm} $	\xi_1 =\xi_1 +l\Delta\xi	$
\STATE \hspace{0.5cm} $	l=l+1 $

\STATE \textbf{Output: }$ \hat{k}_m, \psi_m (z,\hat{k}_m )$ with $m=1,2,\cdots,M$

\end{algorithmic}
\label{alg2}
\end{algorithm}

\section{Simulation }
\begin{figure}[!t]
\centering
\includegraphics[width=3.5in]{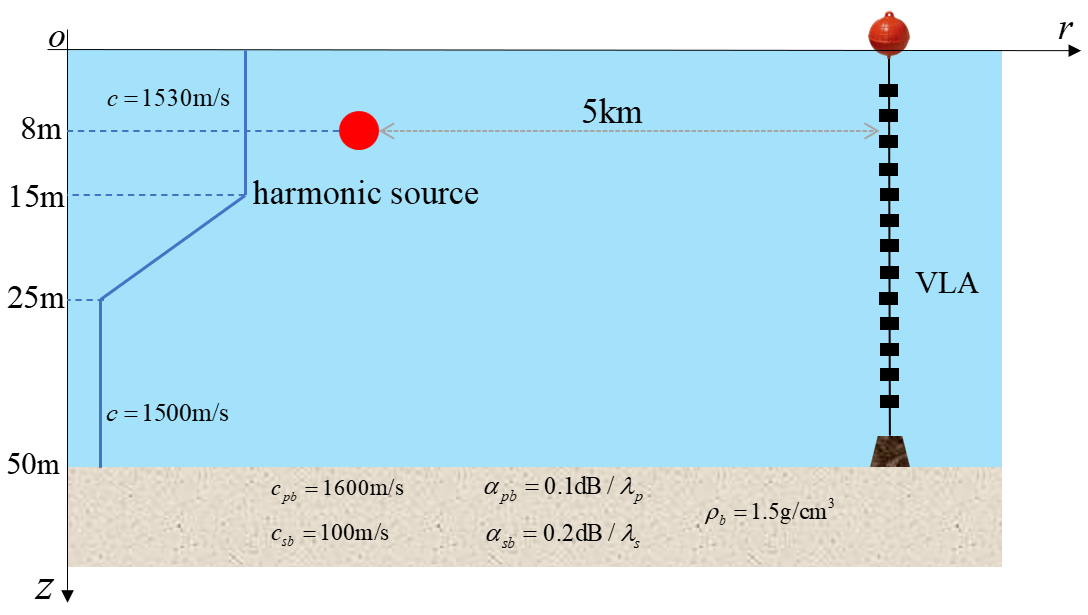}
\caption{Schematic diagram for the simulation environment used in this paper. A 500 Hz harmonic source positioned at 8 m depth maintains a 5 km horizontal separation from the  VLA.
$c_{pb}, c_{sb}, \alpha_{pb}, \alpha_{sb}, \lambda_p, \lambda_s,$ and $\rho_b$ represent
the seabed P-wave sound speed, S-wave sound speed, P-wave attenuation coefficient, S-wave attenuation coefficient,
 P-wave wavelength, S-wave wavelength and density, respectively.}
\label{fig_1}
\end{figure}
\begin{figure}[!t]
\centering
\includegraphics[width=3.5in]{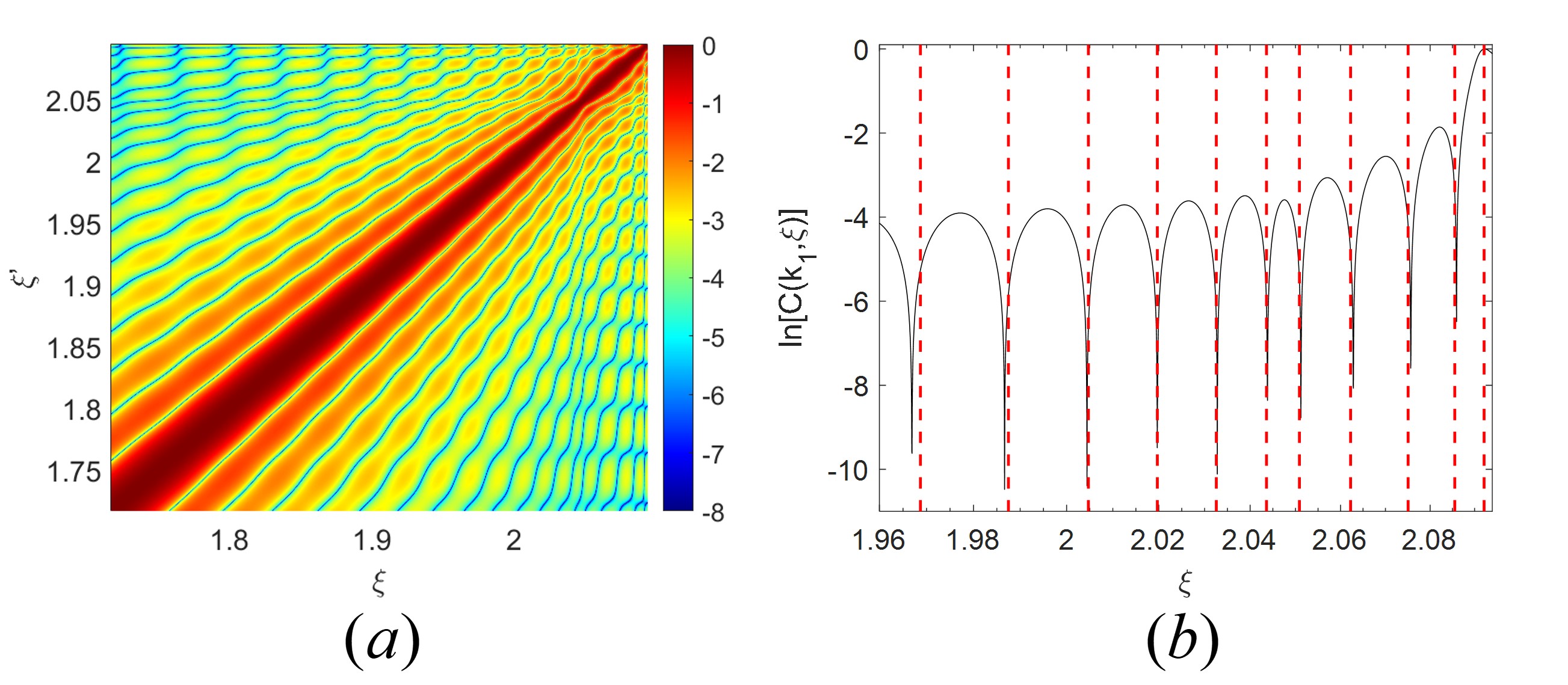}
\caption{(a) $\mathrm{ln}[C(\xi,\xi')]$ and (b) $\mathrm{ln}[C(k_1,\xi)]$ with red dashed lines being the MWs calculated by Kraken.} 
\label{figs2}
\end{figure}

This section presents a comprehensive performance analysis of the OCMS method. The simulation environment, illustrated in Fig. 1, features a linear sound speed profile decreasing from 1530 m/s to 1500 m/s between 15-25 m depth. A 500 Hz omnidirectional harmonic source is positioned at 8 m depth with a 5 km horizontal separation from the VLA. The VLA comprises 50 elements uniformly spaced at 1 m intervals from 1 m to 50 m depth. 
 To validate the performance of the OCMS method, we conduct 50 independent Monte Carlo trials by adding independent complex Gaussian noise to each VLA element while maintaining a system-level signal-to-noise ratio (SNR) of 30 dB, where SNR is defined as: \begin{equation}
    \mathrm{SNR}=10\lg \frac{\|\mathbf{p}\|_2}{\|\mathbf{n}\|_2} 
  \end{equation}
with $\mathbf{n}$ denoting the noise vector and $\mathbf{p}$ representing the signal vector.
For any given MW $\xi$, the corresponding MDF $\psi(z,\xi)$ can be computed using Eq (\ref{eq2}) with a discretization step $h=0.1$ m. Fig. \ref{figs2} (a) displays $\mathrm{ln}[C(\xi,\xi')]$ with
\[
C(\xi,\xi')=\langle \psi(z,\xi), \psi(z,\xi') \rangle.
\]
The results in Fig. \ref{figs2} (a) reveal that for each modal wavenumber $\xi$, multiple local minima exist in $\mathrm{ln}[C(\xi,\xi')]$. $\Phi(z,\xi)$ can be constructed by combining $\xi$ with all  $\xi'$ values corresponding to these minima.
In the following simulations, $M$ in Eq. (\ref{eq7}) equals 18, \textcolor{red}{which is greater than the number of propagating modes.} Fig. \ref{figs2} (b) displays both the $\mathrm{ln}[C(k_1,\xi)]$ and the corresponding MWs calculated by Kraken, showing that all MWs except the largest one are located in the vicinity of local minima of $\mathrm{ln}[C(k_1,\xi)]$.  
Fig. \ref{figs2} (a) and (b) collectively demonstrate the fundamental principle of the OCMS method: the determination of the normal mode parameters through optimization of a single key parameter.

\begin{figure}[!t]
\centering
\includegraphics[width=3.5in]{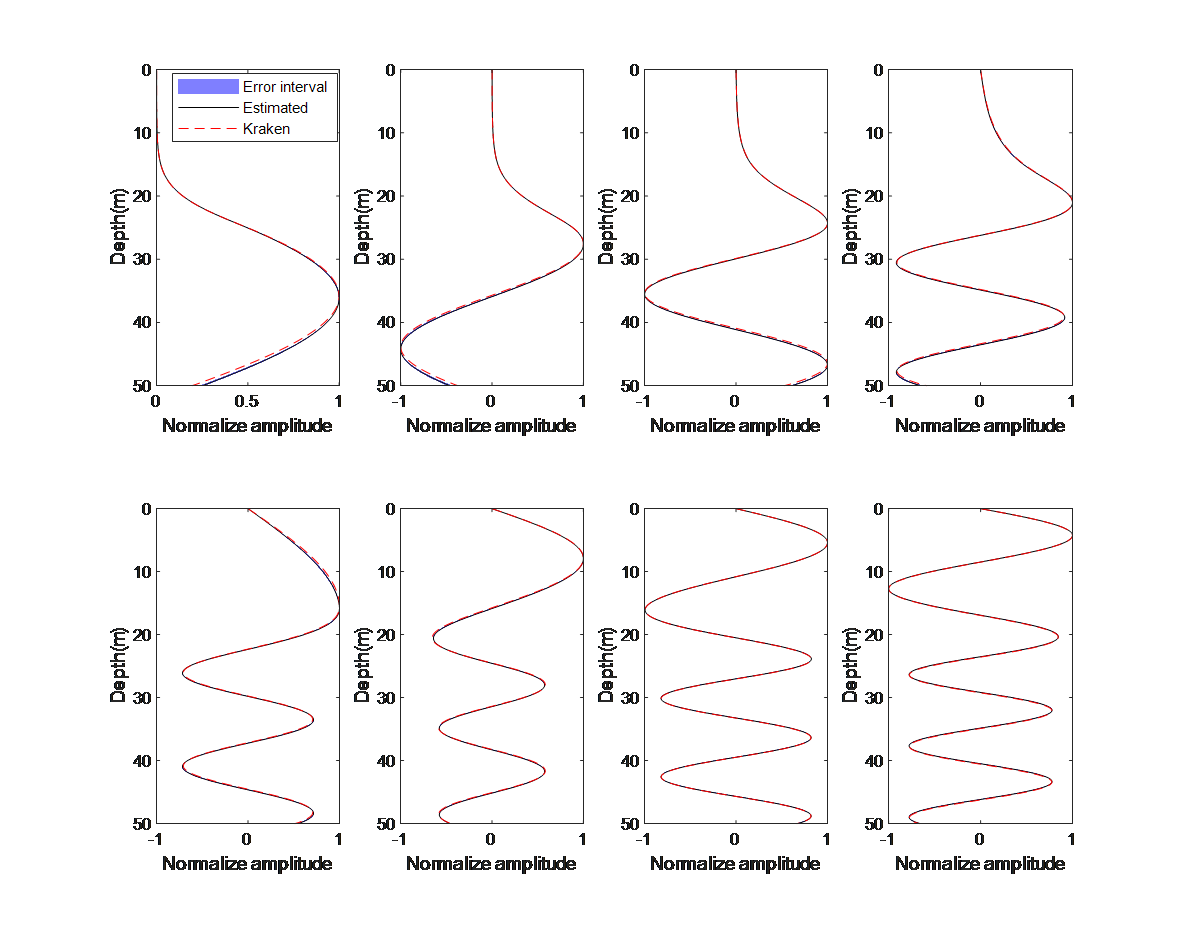}
\caption{Average MDFs estimated from 50 Monte Carlo trials using the OCMS method, with corresponding Kraken-computed MDFs shown as red dashed curves for reference.  }
\label{fig_2}
\end{figure}

Table \ref{tab:table1} and Fig. \ref{fig_2} \textcolor{red}{give} MWs and MDFs estimation results respectively based on OCMS method.
In Table \ref{tab:table1}, $k_m $ represents the reference MW calculated with the Kraken code \cite{ref40},
\begin{equation}
\overline{\hat{k}}_{m}=\frac{1}{50} \sum_{l=1}^{50} \hat{k}_{m}^{l}
\end{equation}
with ${\hat{k}}_{m}^l$ being the $m\mathrm{th}$ normal mode MW estimated by the $l\mathrm{th}$ Monte Carlo realizations,
\begin{equation}
\label{eq12}
\Delta k_m=\frac{|\overline{\hat{k}}_{m}-k_m |}{k_m}
\end{equation}
 measures the relative difference between $k_m$ and $\overline{\hat{k}}_{m}$,
 and $\sigma(k_m )$ is the standard deviation of $\hat{k}_{m}^l$.  Table \ref{tab:table1} reveals that both $\Delta k_m$ and $\sigma(k_m )$ maintain magnitudes on the order of $10^{-4}$, demonstrating the effectiveness of the OCMS method for MW estimation.
Fig. \ref{fig_2} displays the averaged MDFs estimation through 50 Monte Carlo trials using the OCMS method, with the corresponding Kraken-calculated MDFs shown for comparison as red dashed curves.
The error interval in Fig. \ref{fig_2} is the interval between $\overline{\psi }_m (z)-\sigma(\psi_m (z))$ and $\overline{\psi }_m (z)+\sigma(\psi_m (z))$, where $\overline{\psi }_m (z)$ represents the average of 50 Monte Carlo-estimated MDFs using the OCMS method, and $\sigma(\psi_m (z))$denotes the corresponding standard deviation.
Fig. \ref{fig_2} demonstrates near-perfect agreement between the MDFs estimated via the OCMS method and those computed by the Kraken code. 
\begin{table}[!t]
\caption{MW estimation results based on the OCMS method.\label{tab:table1}}
\centering
\begin{tabular}{|c||c||c||c||c|}
\hline
Mode number &	$k_m$	& $\hat{k}_m$ &	$\Delta k_m$	& $\sigma(k_m )$\\
\hline
1	&2.092152	 	&2.092251	&$9.7\times10^{-5}$	&$2.6\times10^{-5}$\\
\hline
2	& $2.085543$	& $	2.085868$	& $	3.2\times10^{-4}$	& $1.0\times10^{-4}$\\
\hline
3	& $2.075066$	& $	2.075556$	& $	4.5\times10^{-4}	$ & $1.3\times10^{-4}$\\
\hline
4	& $2.062222$	& $	2.062723$	& $	5.0\times10^{-4}$	& $1.8\times10^{-4}$\\
\hline
5	& $2.050872$	& $	2.051117$	& $	2.4\times10^{-4}$	& $1.0\times10^{-4}$\\
\hline
6	& $2.043594$	& $	2.043752$	& $	1.6\times10^{-4}$	& $1.0\times10^{-4}$\\
\hline
7	& $2.032576$	& $	2.032665$	& $	8.9\times10^{-5}$	& $1.6\times10^{-4}$\\
\hline
8	& $2.019683$	& $	2.019539$	& $	1.4\times10^{-4}$	& $1.6\times10^{-4}$\\
\hline
9	& $2.004699$	& $	2.004251$	& $	4.4\times10^{-4}$	& $2.1\times10^{-4}$\\
\hline
10	& $1.987477$	& $	1.986498$	& $	9.8\times10^{-4}$ & $	2.1\times10^{-4}$\\
\hline
11	& $1.968707$	& $	1.966743$	& $	2.0\times10^{-3}$ & 	$2.3\times10^{-4}$\\
\hline
\end{tabular}
\end{table}

\par 	\textcolor{red}{The subsequent analysis examines the OCMS method under varying SNR conditions, different VLA apertures, distinct VLA element counts, the effect of VLA tilt and SSP uncertainty.}
\subsection{Varying SNR conditions}
\begin{figure*}[!t]
\centering
\subfloat[]{\includegraphics[width=2.5in]{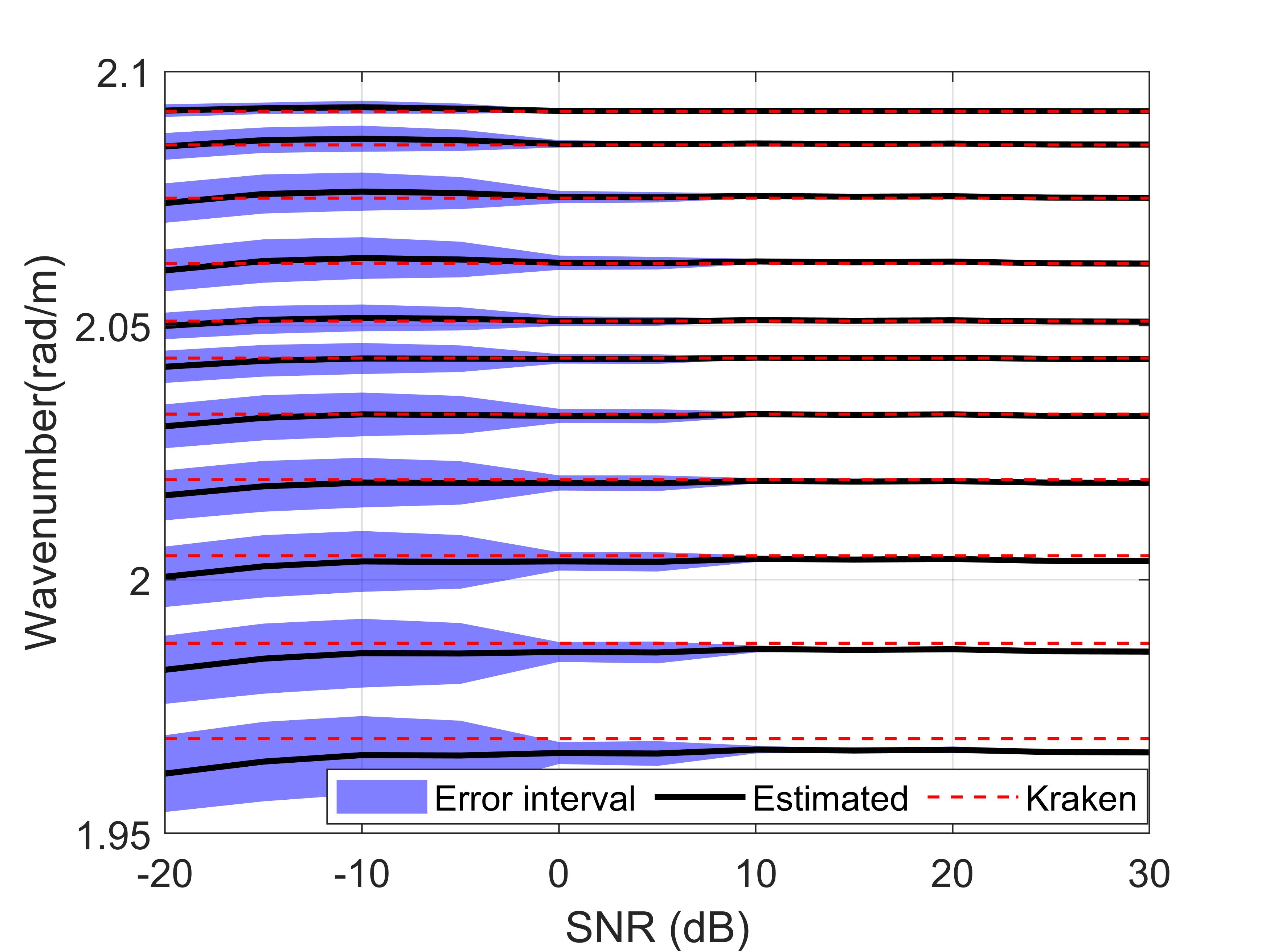}%
\label{(a)}}
\hfil
\subfloat[]{\includegraphics[width=2.5in]{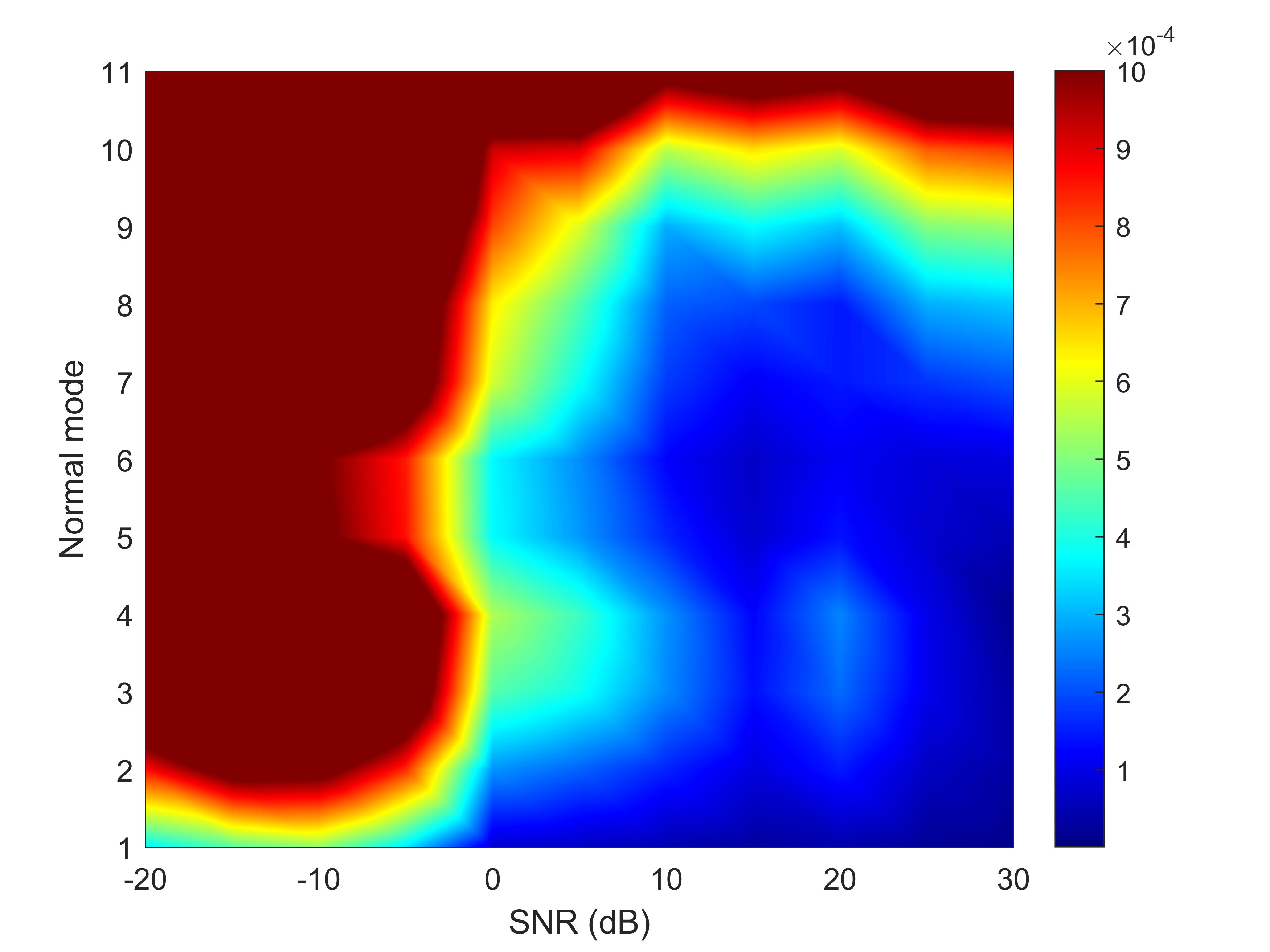}%
\label{(b)}}
\caption{(a) Averaged MW estimates from all Monte Carlo trials using the OCMS method under varying SNR conditions, compared with Kraken-computed reference MWs . 
(b) Corresponding relative errors in MW estimation.}
\label{fig_3}
\end{figure*}

This section evaluates the OCMS method performance across varying SNR conditions, while maintaining a fixed VLA configuration with 50 m aperture spanning the entire water column.
For simulation simplicity, \textcolor{red}{ additive white Gaussian noise }is introduced to the acoustic signals, with SNR varying from -20 dB to 30 dB.
The uncertainty and the relative error of the OCMS method are evaluated through 50 independent Monte Carlo trials,
with corresponding results \textcolor{red}{ presented} in Fig.\ref{fig_3} (a) and (b). 
Fig.\ref{fig_3} (a) shows that the uncertainty of MW estimation decreases as the SNR increases, while Fig.\ref{fig_3} (b) confirms this trend through the corresponding reduction in relative error across all estimated MWs. Both results demonstrate the improved performance of the OCMS method at higher SNR levels. 
The results presented in Fig. \ref{fig_3} demonstrate that the OCMS method achieves reliable performance in our simulations when the SNR exceeds 10 dB. 

\subsection{Different VLA apertures}
\begin{figure*}[!t]
\centering
\subfloat[]{\includegraphics[width=2.5in]{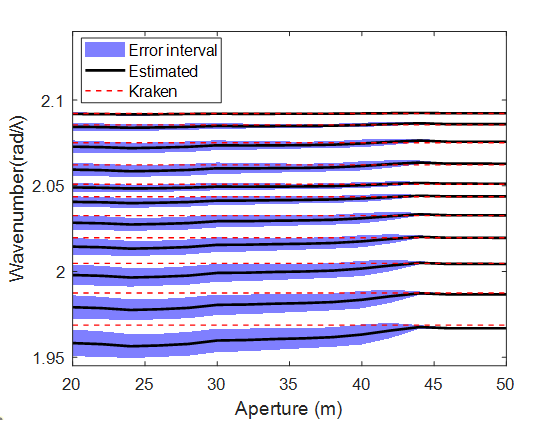}%
\label{(a)}}
\hfil
\subfloat[]{\includegraphics[width=2.5in]{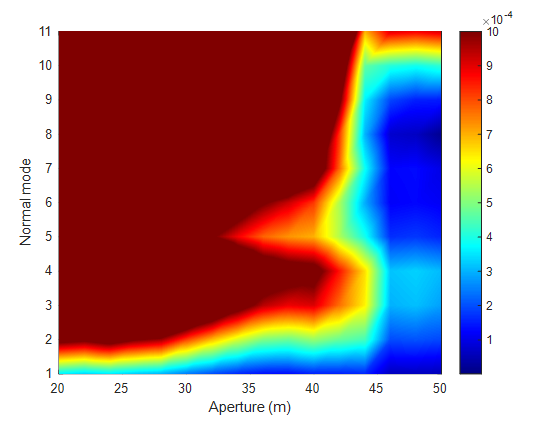}%
\label{(b)}}
\caption{(a) Averaged MW estimates from all Monte Carlo trials using the OCMS method for different VLA apertures, compared with Kraken-computed reference MWs. (b) Corresponding relative errors in MW estimation.}
\label{fig_4}
\end{figure*}

This analysis maintains a constant 30 dB SNR while varying the VLA aperture from 20 to 50 m in 1-m increments, with fixed 1 m element spacing. For apertures less than 50 m, the deployment depth of the VLA within the water column can be adjusted. \textcolor{red}{To evaluate the impact of VLA aperture on the OCMS method performance, we examine configurations in which the VLA aperture is less than 50 m. For each aperture size $A$, we consider deployment scenarios with the first array element positioned at depths from 1 m to$ (50 - A + 1)$ m below the sea surface, incrementing by 1 m. The OCMS method performance is assessed through 50 independent Monte Carlo trials for each deployment configuration.}
Fig. \ref{fig_4} (a) displays the averaged MW estimates (denoted as $\overline{\hat{k}}_{m}$) obtained from all Monte Carlo trials using the OCMS method in different VLA apertures. For reference, the corresponding MWs calculated by the Kraken code are shown as red dashed curves. 
Fig. \ref{fig_4} (b) presents the relative differences between the averaged MW estimates $\overline{\hat{k}}_{m}$ and the reference values $k_m$ for modes $m = 1$ through 11 in various apertures. Analysis of Fig. \ref{fig_4} (a) and (b) reveals two findings: for VLA apertures exceeding 43 m, all 11 modes exhibit relative estimation errors below $9\times10^{-4}$, while apertures beyond 35 m maintain this error threshold for the first three modes.

\subsection{ Distinct VLA element counts}
\begin{figure*}[!t]
\centering
\subfloat[]{\includegraphics[width=2.5in]{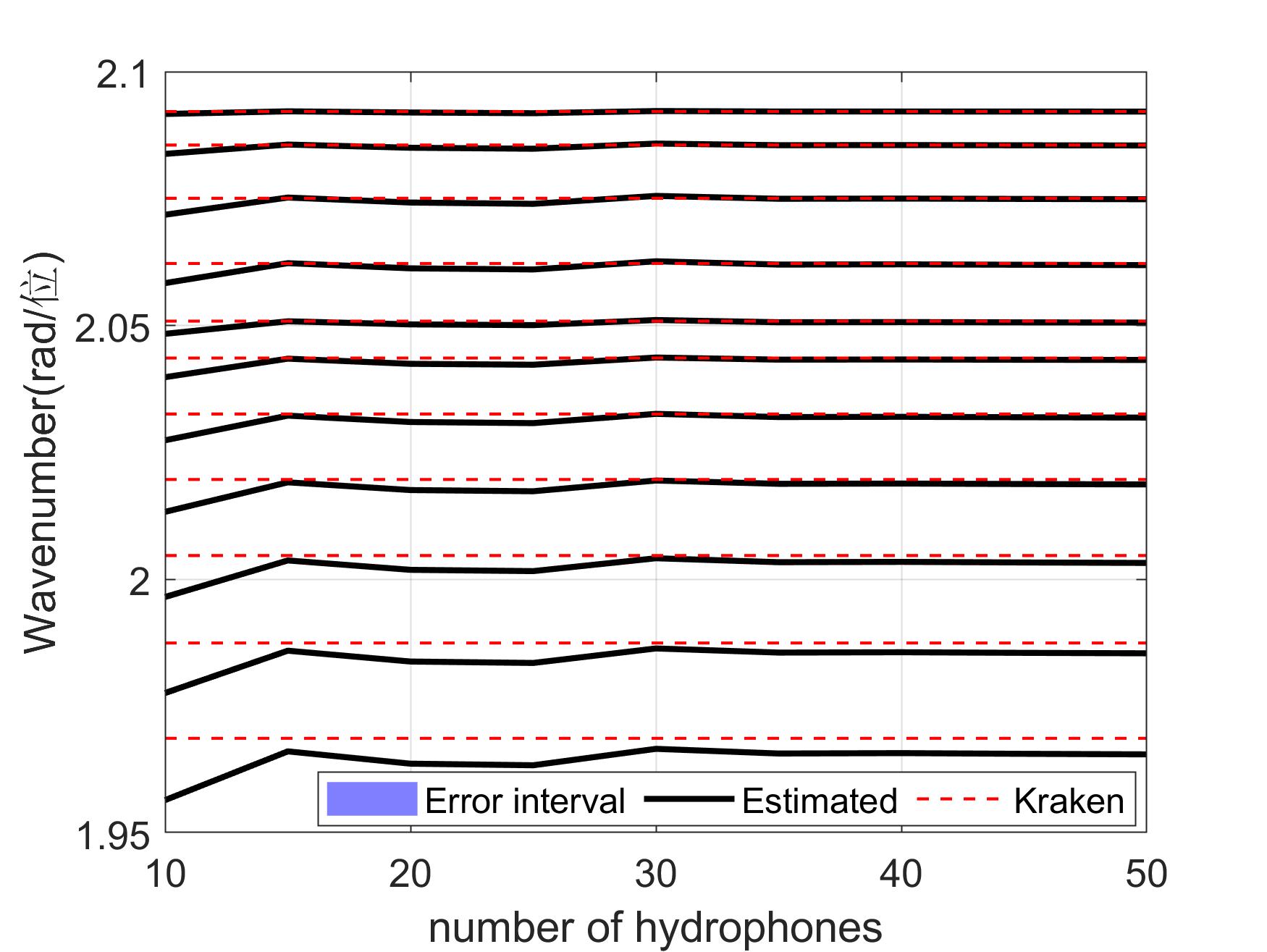}%
\label{(a)}}
\hfil
\subfloat[]{\includegraphics[width=2.5in]{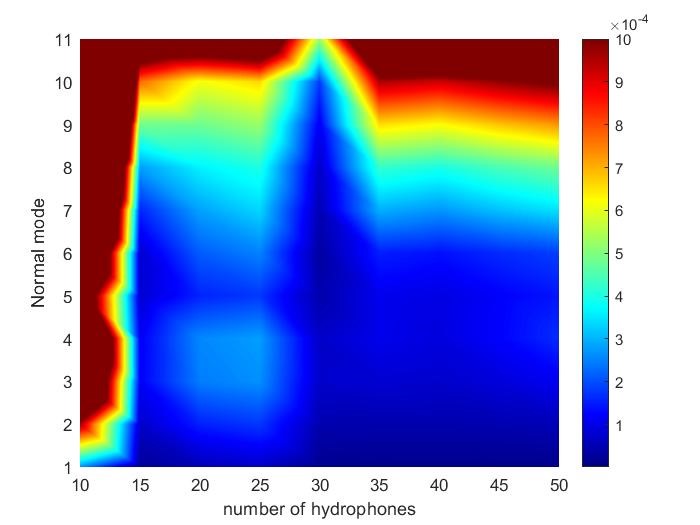}%
\label{(b)}}
\caption{(a) Averaged MW estimates from all Monte Carlo trials using the OCMS method for distinct VLA element counts, compared with Kraken-computed reference MWs.
(b) Corresponding relative errors in MW estimation.}
\label{fig_5}
\end{figure*}

This subsection evaluates the influence of hydrophone quantity in the VLA on the OCMS method performance. The VLA with 50 m aperture fully spans the water column while maintaining a constant 25 dB SNR. 
\textcolor{red}{The VLA configuration employs 10 to 50 hydrophones, with inter-element spacing varying inversely from 5 m to 1 m as the number of elements increases.} 
Fig. \ref{fig_5} (a) shows that the combination of high SNR and full aperture deployment yields low uncertainty in MW estimation, although significant deviations occur when the VLA contains too few hydrophones. The corresponding relative errors are presented in Fig. \ref{fig_5} (b), which shows that for VLA configurations with more than 15 hydrophones, the first nine modes maintain relative estimation errors below $ 9 \times 10^{-4}$. 

\subsection{ The Effect of  VLA tilt}
\textcolor{red}{This subsection evaluates the effect of VLA tilt on the performance of the OCMS method. Due to the effects of ocean currents and waves, the Vertical Line Array (VLA) deployed in the ocean is often tilted. Generally, VLA tilt causes variations in the array element data, including amplitude and phase, with the phase shift being particularly significant. In the simulations conducted in this section, the VLA spans the entire water column from 1 meter to 50 meters, with a 1-meter spacing between elements, and SNR is 25 dB. When the VLA tilts at an angle $\theta$ , the actual positions of each array element can be calculated using the following formula.
\begin{equation}
\label{eqVLAtitl}
z_n'=z_n\mathrm{cos}\theta+H(1-\mathrm{cos}\theta),
r_n'=r_0+(H-z_n)\mathrm{sin}\theta,
\end{equation}
 where $z_n$ is the designed depth of the $n$-th array element, and  $r_0$ is the designed range between the VLA and sound source. According to the equation, the actual position of the $n$-th  element of the array is related to the tilt angle of the VLA. 
 } 
 \par
\begin{figure*}[!t]
\centering
\subfloat[]{\includegraphics[width=2.5in]{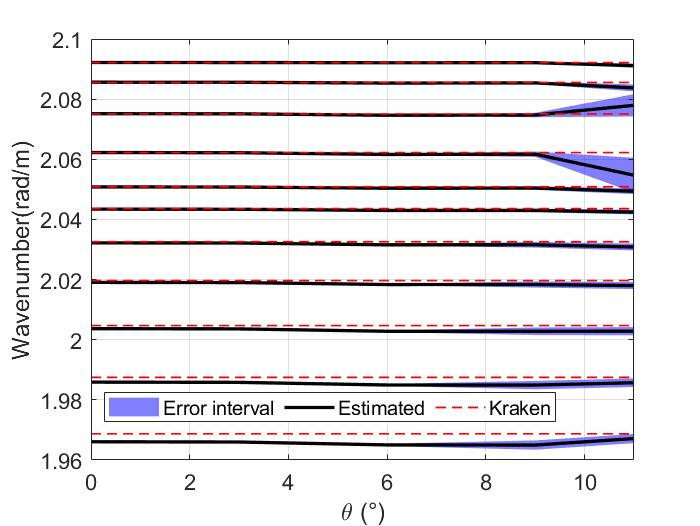}%
\label{(a)}}
\hfil
\subfloat[]{\includegraphics[width=2.5in]{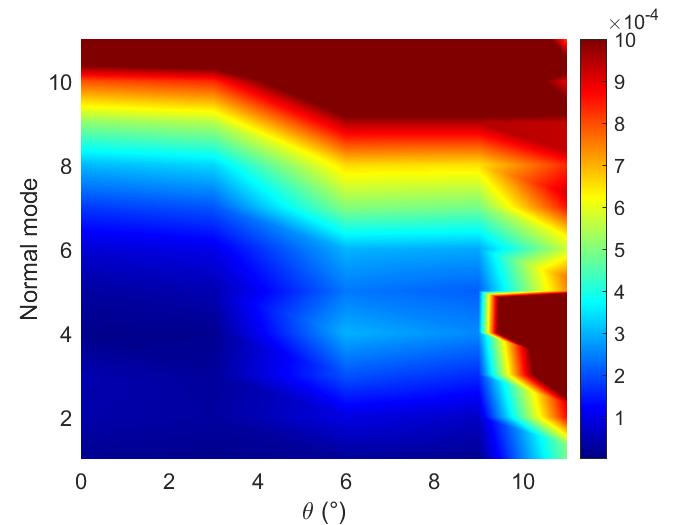}%
\label{(b)}}
\caption{(a) Averaged MW estimates from all Monte Carlo trials using the OCMS method for the VLA tilt , compared with Kraken-computed reference MWs.
(b) Corresponding relative errors in MW estimation.}
\label{fig_vla}
\end{figure*}

 \textcolor{red}{It is assumed that the acoustic field data is measured on the actual (tilted) array.However,  the VLA tilt information is not provided to the OCMS algorithm, thus, the algorithm can only use the designed depths of the VLA - implicitly assuming that the VLA is not tilted. When the tilt angle $\theta$ varies from $0^\degree$ to $11^\degree$, the OCMS estimation results is presented in Fig. \ref{fig_vla}. The lower-order normal modes are more robust to the VLA tilt compared with the higher-order modes. When the tilt angle is small ($\theta< 5\degree$), the OCMS method maintains high precision and accuracy, with the relative estimation errors and standard deviations both on the order of $10^{-4}$). The performance of the OCMS method degrades with increasing VLA tilt angle. And simulation results indicate that estimation accuracy deteriorates significantly beyond $10 \degree$, rendering the output unreliable for practical applications.}
 
\subsection{ The Effect of  SSP uncertainty }
\textcolor{red}{
This subsection evaluates the effect of sound speed profile (SSP) uncertainty on the OCMS method performance. In the simulations conducted in this section, the VLA spans the entire water column from 1 meter to 50 meters, with a 1-meter spacing between elements, and SNR is 25 dB. SSP uncertainty presents another practical challenge for OCMS implementation. Measurement uncertainties inevitably arise from both instrumental errors and dynamic oceanographic processes, including sensor calibration drift, spatiotemporal variability of water masses,internal wave-induced perturbations,etc. The sound velocity uncertainty in the water is added by the following equation,
\begin{equation}
\label{eq_ssp}
c(z)=c_0(z)+\eta\alpha,
\end{equation}
 where $c_0(z)$ is the background sound speed profile of the water column, while $\eta$ is a uniformly distributed random variable over the interval [-1,1], and $\alpha$ is a sound velocity variable ranging from 0.1 $m / s$ to 1 $m / s$. 
 In this part, the acoustic field was calculated under the background sound speed profile $c_0(z)$ , while the OCMS algorithm corresponds to the perturbed SSP $c(z)$ . As the SSP uncertainty increases from 0.1 m/s to 1 m/s, the OCMS results are shown in Fig. \ref{fig_ssp}. Although there are clear deviations in the estimated modal wavenumbers (on the order of $10^{-3}$), the standard deviation of the estimation (not shown in the paper) is quite small, on the order of $10^{-4}$.The results show that the lower-order normal modes are significantly less affected than the higher-order modes, and the wavenumber estimation error increases with increasing SSP uncertainty. Numerical simulations demonstrate that the OCMS algorithm retains acceptable estimation precision for SSP uncertainties less than 1 m/s.
 \begin{figure*}[!t]
\centering
\subfloat[]{\includegraphics[width=2.5in]{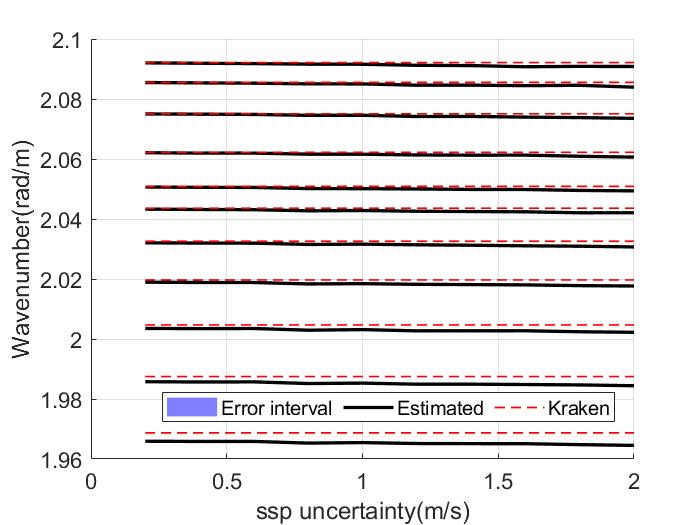}%
\label{(a)}}
\hfil
\subfloat[]{\includegraphics[width=2.5in]{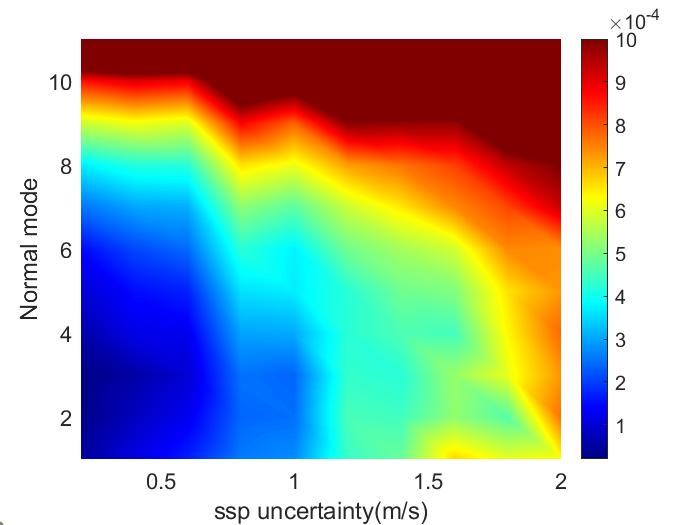}%
\label{(b)}}
\caption{(a) Averaged MW estimates from all Monte Carlo trials using the OCMS method for the SSP uncertainty , compared with Kraken-computed reference MWs.
(b) Corresponding relative errors in MW estimation.}
\label{fig_ssp}
\end{figure*}
}
\section{Experiment}
\begin{figure*}[!t]
\centering
\subfloat[]{\includegraphics[width=2.5in]{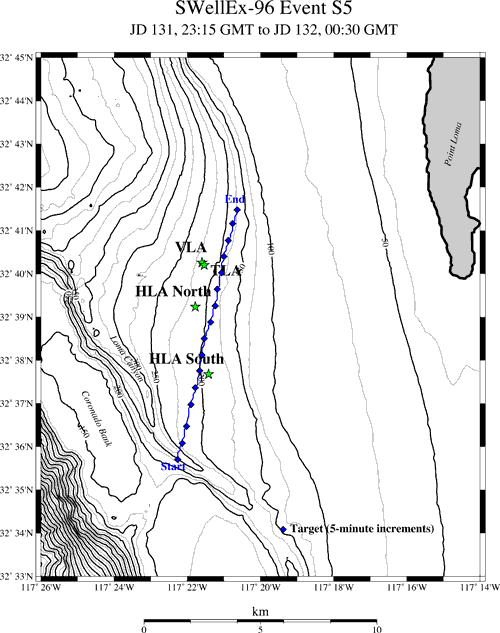}%
\label{(a)}}
\hfil
\subfloat[]{\includegraphics[width=2.5in]{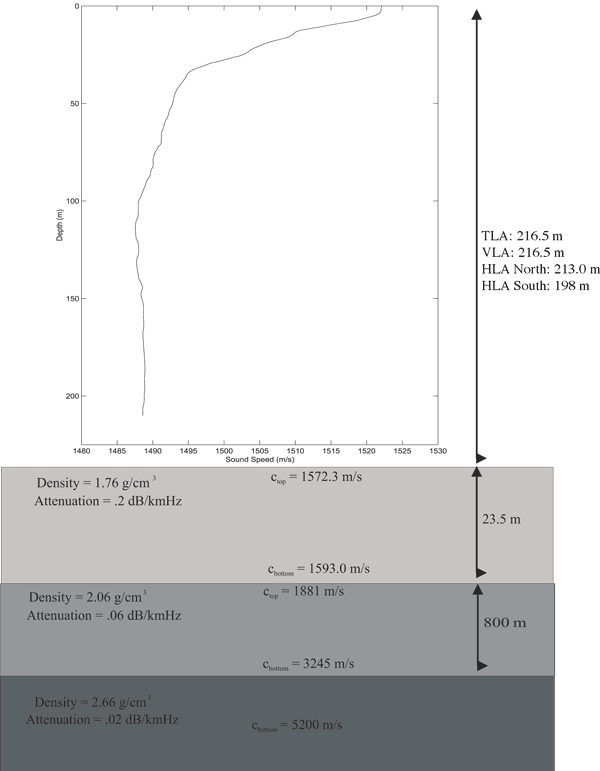}%
\label{(b)}}
\caption{(a) S5 event sound ship trajectory and the location of arrays deployment.
(b) SWellEx-96 experiment sound speed profiles of water column and geoacoustic parameters.}
\label{fig_6}
\end{figure*}

\begin{figure}[!t]
\centering
\includegraphics[width=3.5in]{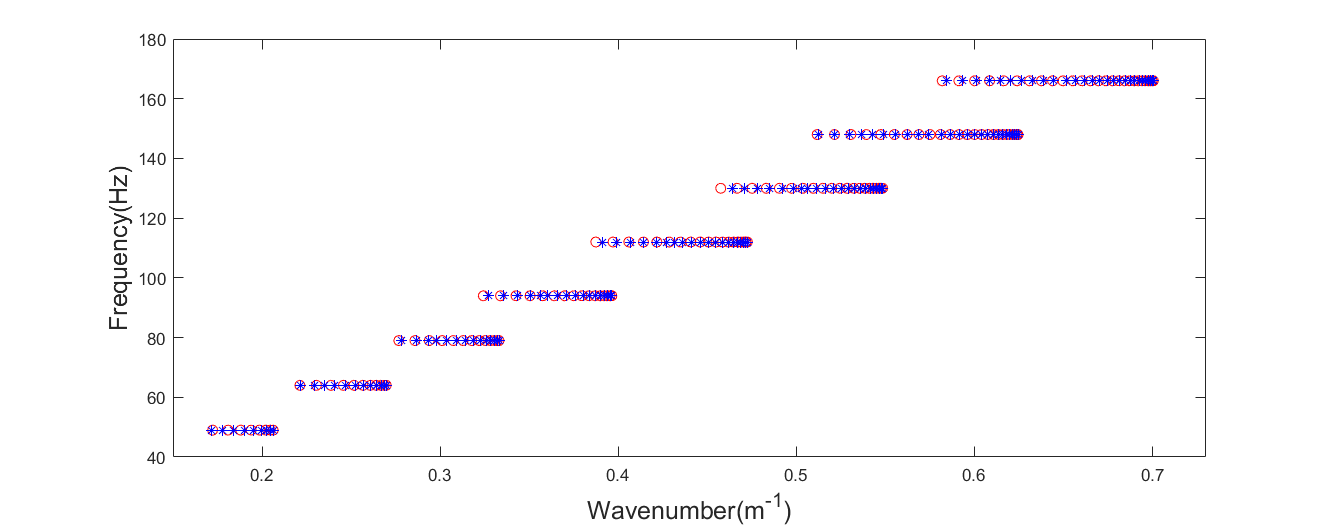}
\caption{Estimated MWs at eight frequencies (49-166 Hz) using the OCMS method (red circles) compared with Kraken-computed values (blue asterisks). }
\label{fig_7}
\end{figure}

This section validates the effectiveness of the OCMS method using the S5 event dataset from the SWellEx-96 experiment \cite{ref41}.
During this event, the source ship (R/V Sproul) towed a deep source and a shallow source from southwest to northeast at 5 knots.
The source trajectory is shown in Fig. \ref{fig_6}(a), with most of the water depths on the trajectory ranging from 180 m to 220 m.
The deep source emitted a series of tones between 49 Hz and 400 Hz, and the shallow source emitted 9 frequencies between 109 Hz and 385 Hz.
During the experiment, a 64-element VLA was deployed at the location shown in Fig. \ref{fig_6}(a), with 21 receiver channels publicly available through the SWellEx-96 database. The array, positioned at 216.5 m water depth, spanned the lower half of the water column. 
Due to the public availability of data from only 21 hydrophones in the VLA, our analysis utilizes exclusively these publicly accessible hydrophone measurements for subsequent processing. 
The OCMS method requires only single-shot processing. We employed data from the 20th to 21st minute of the S5 event, combined with Fourier analysis, to prepare a high-SNR snapshot. Fig. \ref{fig_7} displays the MWs estimated by the OCMS method for nine discrete tones (49 Hz, 64 Hz, 79 Hz, 94 Hz, 112 Hz, 130 Hz, 148 Hz, 166 Hz) from the deep source, with identified modes marked by red circles. These OCMS estimates are compared against benchmark MWs (blue asterisks) computed by the Kraken code, where the environmental parameter configuration exactly matches that shown in Fig. \ref{fig_6}(b).  Fig. \ref{fig_7} demonstrates strong agreement between the MWs estimated by the OCMS method and those computed by the Kraken code. 

\begin{table*}[htbp]
  \centering
  \caption{Relative errors between OCMS-estimated and Kraken-calculated MWs for the first three modes across eight frequencies (49-166 Hz). }
    \begin{tabular}{|c||ccc||ccc||ccc|}
    \hline
    \multicolumn{1}{|c||}{\multirow{2}{*}{$f$}} & \multicolumn{3}{c||}{1st mode} & \multicolumn{3}{c||}{2nd mode}
    & \multicolumn{3}{c|}{3rd mode} \\
\cline{2-10}
\multicolumn{1}{|c||}{}& {$k_1$} & {$\hat{k_1}$}& {$e(k_1)$} & {$k_2$} & {$\hat{k_2}$} & {$e(k_2)$} & {$k_3$} & {$\hat{k_3}$} & {$e(k_3)$} \\
\hline
49    & 0.20635 & 0.20626 & $4.4\times10^{-4}$ & 0.20497 & 0.20464 & $1.6\times10^{-3}$ & 0.20277 & 0.20212 & $3.2\times10^{-4}$ \\
    \hline
    64    & 0.26973 & 0.26972 & $4.8\times10^{-5}$ & 0.26857 & 0.26854 & $1.2\times10^{-4}$ & 0.26681 & 0.26674 & $2.7\times10^{-4} $\\
    \hline
    79    & 0.33309 & 0.33306 & $9.3\times10^{-5}$ & 0.33206 & 0.33194 & $3.6\times10^{-4}$ & 0.33057 & 0.33036 & $6.3\times10^{-4}$ \\
    \hline
    94    & 0.39644 & 0.39642 & $6.1\times10^{-5}$ & 0.3955 & 0.39544 & $1.5\times10^{-4}$ & 0.39418 & 0.39408 & $2.5\times10^{-4}$ \\
    \hline
    112   & 0.47245 & 0.47248 & $5.9\times10^{-5}$ & 0.47159 & 0.47171 & $2.5\times10^{-4}$ & 0.4704 & 0.47054 & $2.9\times10^{-4}$ \\
    \hline
    130   & 0.54845 & 0.54843 & $4.2\times10^{-5}$ & 0.54766 & 0.54759 & $1.2\times10^{-4}$ & 0.54655 & 0.54646 & $1.6\times10^{-4} $\\
    \hline
    148   & 0.62445 & 0.62448 & $5.0\times10^{-5}$ & 0.62372 & 0.62384 & $1.9\times10^{-4}$ & 0.62266 & 0.62279 & $2.1\times10^{-4}$ \\
    \hline
    166   & 0.70039 & 0.70048 & $1.3\times10^{-4}$ & 0.6997 & 0.69975 & $7.0\times10^{-5}$ & 0.69868 & 0.69871 & $6.2\times10^{-4}$ \\
    \hline
    \end{tabular}%
  \label{tab:addlabe2}%
\end{table*}%

Table \ref{tab:addlabe2} provides quantitative validation of the OCMS method by comparing its estimated modal wavenumbers for the first three modes at eight frequencies against benchmark values computed by the Kraken code, along with their corresponding relative errors. 
The relative error for each MW  is defined as
\begin{equation}
   e\left(k_{m}\right)=\left|\frac{k_{m}-\hat{k}_{\mathrm{m}}}{k_{\mathrm{m}}}\right| 
\end{equation}
where $\hat{k}_{\mathrm{m}}$ is the estimated $m\mathrm{th}$ MW by the OCMS method. 
The same analysis procedure was  \textcolor{red}{repeated} to nine frequency tones emitted by the shallow source, with the first three MWs presented in Table \ref{tab:addlabe3}. Both Tables \ref{tab:addlabe2} and \ref{tab:addlabe3} show that the relative errors between estimated and calculated MWs for the first three modes maintain magnitudes on the order of $10^{-4}$, further validating the effectiveness of the OCMS method. 
Tables \ref{tab:addlabe2} and \ref{tab:addlabe3}  reveal an interesting observation: the relative errors for the first three modes decrease systematically with increasing frequency. This trend correlates with reduced seabed penetration capability of the MDFs at higher frequencies. The observed improvement in OCMS performance stems from better satisfaction of the orthogonality assumption when modal energy becomes increasingly confined within the water column. 

\begin{table*}[htbp]
  \centering
\caption{Relative errors between OCMS-estimated and reference MWs for the first three modes across nine frequencies (shallow source configuration). }
    \begin{tabular}{|c||ccc||ccc||ccc|}
    \hline
    \multicolumn{1}{|c||}{\multirow{2}{*}{$f$}} & \multicolumn{3}{c||}{1st mode} & \multicolumn{3}{c||}{2nd mode}
    & \multicolumn{3}{c|}{3rd mode} \\
\cline{2-10}
\multicolumn{1}{|c||}{}& {$k_1$} & {$\hat{k_1}$}& {$e(k_1)$} & {$k_2$} & {$\hat{k_2}$} & {$e(k_2)$} & {$k_3$} & {$\hat{k_3}$} & {$e(k_3)$} \\
\hline
    109   & 0.45978 & 0.45991 & 2.9$\times 10^{-4}$ & 0.45891 & 0.45924 & 7.1$\times 10^{-4}$ & 0.4577 & 0.45805 & 7.5$\times 10^{-4}$ \\
    \hline
    127   & 0.53578 & 0.53591 & 2.4$\times 10^{-4}$ & 0.53498 & 0.53531 & 6.1$\times 10^{-4}$ & 0.53386 & 0.53419 & 6.1$\times 10^{-4}$ \\
    \hline
    145   & 0.61178 & 0.61192 & 2.3$\times 10^{-4}$ & 0.61104 & 0.61139 & 5.7$\times 10^{-4}$ & 0.60998 & 0.61031 & 5.4$\times 10^{-4}$ \\
    \hline
    163   & 0.68778 & 0.68795 & 2.5$\times 10^{-4}$ & 0.68709 & 0.98746 & 5.4$\times 10^{-4}$ & 0.68606 & 0.68641 & 5.2$\times 10^{-4}$ \\
    \hline
    198   & 0.83555 & 0.83561 & 8.0$\times 10^{-5}$ & 0.83494 & 0.83517 & 2.7$\times 10^{-4}$ & 0.83395 & 0.83419 & 2.9$\times 10^{-4}$ \\
    \hline
    232   & 0.97909 & 0.9791 & 1.1$\times 10^{-5}$ & 0.97855 & 0.97863 & 8.0$\times 10^{-5}$ & 0.97758 & 0.97766 & 8.1$\times 10^{-5}$ \\
    \hline
    280   & 1.18173 & 1.1817 & 3.0$\times 10^{-5}$ & 1.18127 & 1.18097 & 2.6$\times 10^{-4}$ & 1.18033 & 1.18002 & 2.7$\times 10^{-4}$ \\
    \hline
    335   & 1.41393 & 1.41401 & 6.0$\times 10^{-5}$ & 1.41354 & 1.41381 & 2.0$\times 10^{-4}$ & 1.41263 & 1.41294 & 2.2$\times 10^{-4}$ \\
    \hline
    385   & 1.62504 & 1.62502 & 1.0$\times 10^{-5}$ & 1.62466 & 1.62435 & 1.9$\times 10^{-4}$ & 1.62378 & 1.62349 & 1.8$\times 10^{-4}$ \\
    \hline
    \end{tabular}%
  \label{tab:addlabe3}%
\end{table*}%

\section{Conclusion}
This paper proposes the orthogonality-constrained modal search (OCMS) method for estimating acoustic normal mode parameters using a vertical linear array (VLA) with single-shot measurements. The method requires only a single-frequency acoustic field measurement at an arbitrary source-receiver distance, eliminating both the need for precise range information and all limitations associated with synthetic aperture horizontal array techniques. 

\par The OCMS method transforms the multidimensional normal mode parameter estimation problem into a one-dimensional modal wavenumber search problem by exploiting the relationship between modal wavenumbers and mode depth functions along with their orthogonality properties, solvable through convex optimization. \textcolor{red}{Simulation results demonstrate the robustness and accuracy of OCMS under varying vertical linear array (VLA) apertures, signal-to-noise ratios (SNR), and element counts. Both VLA tilt and SSP uncertainty introduce systematic biases in OCMS estimates, with higher-order normal modes exhibiting greater susceptibility to these perturbations. Simulation results demonstrate that the OCMS method maintains robust performance as long as the VLA tilt angle is small than  $5\degree$ and the SSP uncertainty is small than 1 m/s.} The effectiveness of the OCMS method is confirmed by experimental data from SwellEx96, showing relative errors of order $10^{-4}$ between OCMS-derived modal wavenumbers and Kraken-calculated benchmarks.

\appendix
In this part, we will clarify that $p(z)$ in Eq. (\ref{eq7}) cannot be represented linearly by a finite number of elements in $\Phi(z,\xi_1)$ when $\xi\neq k_1$. Assume
\[
p(z)=\sum_{m=1}^{M} a_m \psi_m (z,k_m)=\sum_{n=1}^{N} b_n \psi_n (z,\xi_n), \;\; M,N \in \mathbb{N}^+,
\]
$\psi_n(z,\xi_n)\in\Phi(z,\xi_1)$.
Then
\begin{equation}
\psi_1 (z,k_1)=\sum_{n=1}^{N} b_n' \psi_n (z,\xi_n).
\label{eqa1}
\end{equation}
Because $\psi_m (z,k_m)$ is the eigenfunction of the operator $\partial_z^2+k^2(z)$ with eigenvalue $k_m^2$. By acting the operator $\partial_z^2+k^2(z)$ on both sides of the above equation, we obtain
\begin{equation}
\psi_1 (z,k_1)=\sum_{n=1}^{N} \frac{\xi_n^2}{k_1^2} b_n' \psi_n (z,\xi_n),
\label{eqa2}
\end{equation}
which is not consistent with Eq. (\ref{eqa1}). So $p(z)$ in Eq. (\ref{eq7}) cannot be represented linearly by a finite number of elements in $\Phi(z,\xi_1)$ when $\xi\neq k_1$.

\newpage

\section{Biography Section}

\begin{IEEEbiography}[{\includegraphics[width=1in,height=1.25in,clip,keepaspectratio]{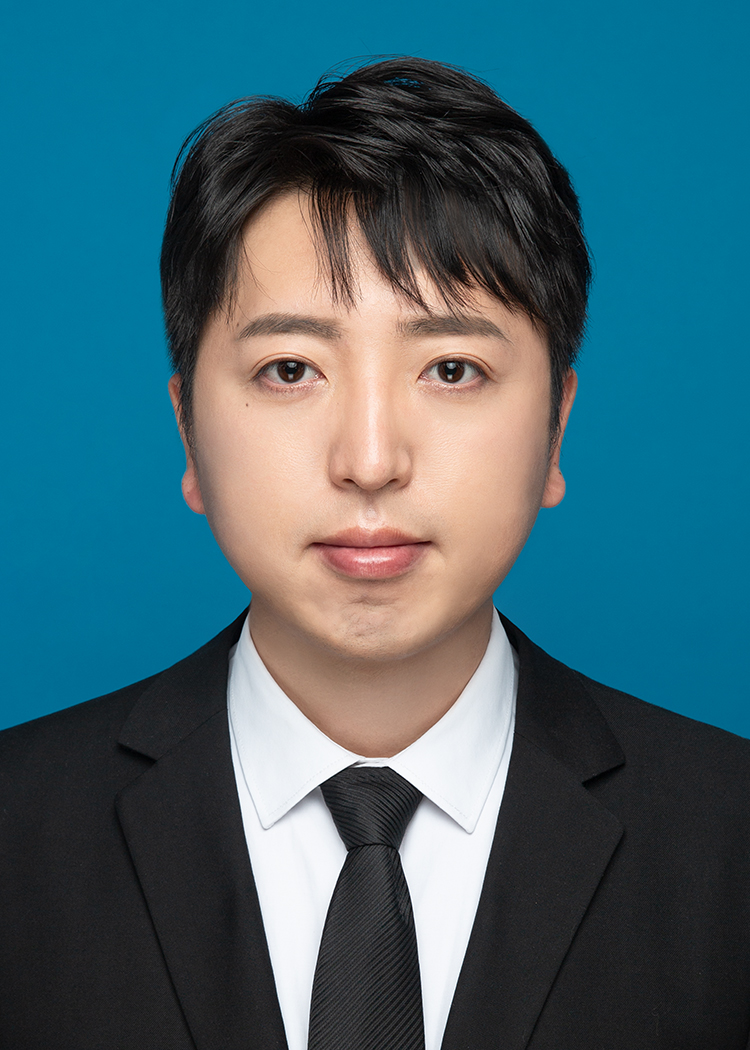}}]{Xiaolei Li}
received the B.S. degree in marine technology and
Ph.D. degrees in marine survey technology from the the Ocean University of China, Qingdao, China, in 2013 and 2019, respectively. He is currently a Lecturer
with the School of Electronic Information Science and Engineering, Ocean University of China. His research interests include ocean acoustics, underwater acoustic signal processing and their applications.
\end{IEEEbiography}

\begin{IEEEbiography}[{\includegraphics[width=1in,height=1.25in,clip,keepaspectratio]{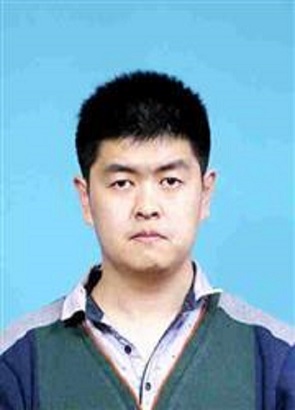}}]{Pengyu Wang} was born in Qingdao, Shandong Province, China in 1989. In 2018, he obtained a PhD in Intelligent Information and Communication from Ocean University of China.
Since 2013, he has successively studied microelectronics, image processing and acoustics. Now working in the Department of Electronics, School of Electronic Information Science and Engineering, Ocean University of China. He has published 7 papers and is currently mainly engaged in signal processing and marine acoustics related works.
\end{IEEEbiography}

\begin{IEEEbiography}
[{\includegraphics[width=1in,height=1.25in,clip,keepaspectratio]{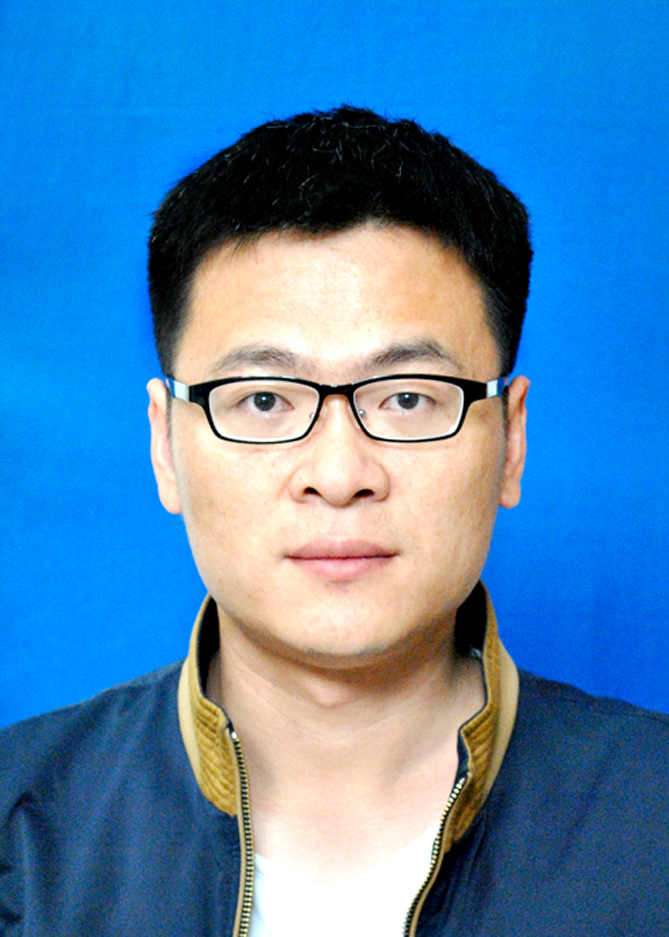}}]
{Wenhua Song} received the B.S. degree in Ocean University of China, Qingdao, China, in 2009 and the M.S. and Ph.D. degree in Acoustics at University of Chinese Academy of Sciences, Beijing, China, in 2014. 
From 2014 to 2017, he was a postdoctoral scholar in Marine Technology in Ocean University of China. Since 2017, He works at the department of Physics in Ocean University of China as a lecturer. His research interest includes the propagation of acoustic waves in dynamic ocean, the geoacoustic inversion of sea bottom sediment, the localization of underwater targets using passive sonar.
\end{IEEEbiography}

\begin{IEEEbiography}
[{\includegraphics[width=1in,height=1.25in,clip,keepaspectratio]{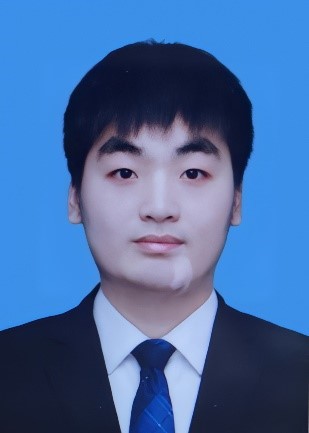}}]
{Yangjin Xu} received the B.S. degree in Acoustic from Anhui Jianzhu University, Hefei, China, in 2022. He is currently pursuing an M.S. degree in Acoustics at the School of Electronic Information Science and Engineering, Ocean University of China. His research interests include underwater acoustic signal processing and underwater target detection.
\end{IEEEbiography}

\begin{IEEEbiography}
[{\includegraphics[width=1in,height=1.25in,clip,keepaspectratio]{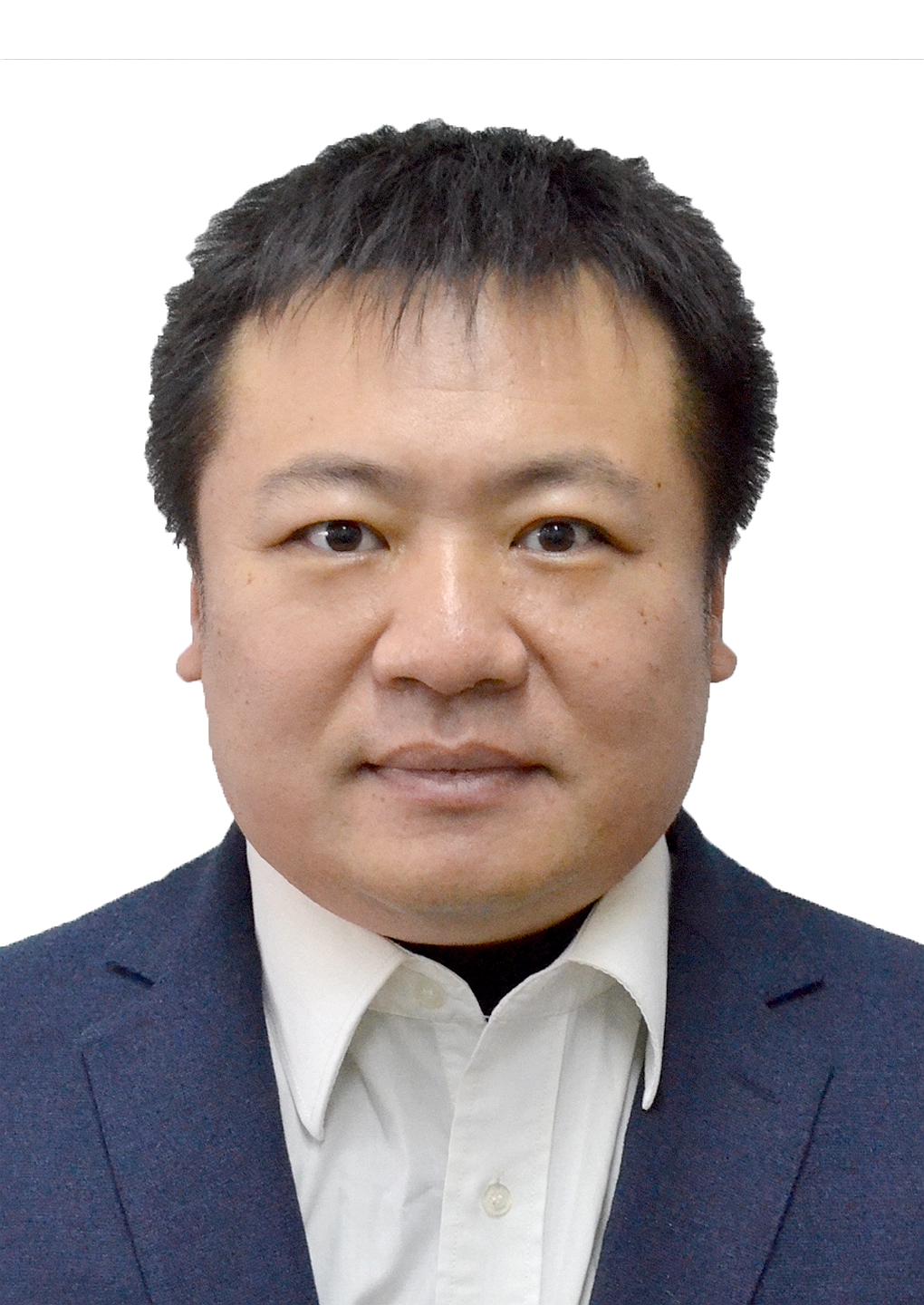}}]
{Wei Gao} received the Ph. D. degree in Marine Survey Technology from the Ocean University of China in 2008. He is a full professor with the School of Electronic Information Science and Engineering, Ocean University of China and vice president of
Shandong Acoustical Society. His research interests include ocean acoustics, underwater acoustic signal processing, cross-application of ocean acoustics, artificial intelligence and big data technology, and joint distribution characteristics of Marine multi-physical field and its applications.
\end{IEEEbiography}

\vfill

\end{document}